\documentclass[sn-mathphys-num]{sn-jnl}

\usepackage{graphicx}%
\usepackage{multirow}%
\usepackage{amsmath,amssymb,amsfonts}%
\usepackage{amsthm}%
\usepackage{mathrsfs}%
\usepackage[title]{appendix}%
\usepackage{xcolor}%
\usepackage{textcomp}%
\usepackage{manyfoot}%
\usepackage{booktabs}%
\usepackage{algorithmicx}%
\usepackage{algpseudocode}%
\usepackage{listings}%

\usepackage{bbm}
\usepackage{tabularx}
\usepackage{makecell}
\usepackage{algorithm2e}

\DeclareMathOperator{\Tr}{Tr}

\usepackage[braket,qm]{qcircuit}
\usepackage{multirow,multicol}

\theoremstyle{thmstyleone}%
\theoremstyle{thmstyletwo}%

\theoremstyle{thmstylethree}%

\begin{document}

\title[Article Title]{Diffusion-Inspired Quantum Noise Mitigation in \\ Parameterized Quantum Circuits}


\author*[1,6]{\fnm{Hoang-Quan} \sur{Nguyen}}\email{hn016@uark.edu,}

\author[1,6]{\fnm{Xuan Bac} \sur{Nguyen}}\email{xnguyen@uark.edu}

\author[2]{\fnm{Samuel Yen-Chi} \sur{Chen}}\email{yen-chi.chen@wellsfargo.com}

\author[3,6]{\fnm{Hugh} \sur{Churchill}}\email{hchurch@uark.edu}

\author[4,6]{\fnm{Nicholas} \sur{Borys}}\email{nicholas.borys@montana.edu}

\author[5]{\fnm{Samee U.} \sur{Khan}}\email{ skhan@ece.msstate.edu}

\author[1,6]{\fnm{Khoa} \sur{Luu}}\email{khoaluu@uark.edu}

\affil*[1]{\orgdiv{Department of Electrical Engineering and Computer Science}, \orgname{University of Arkansas}, \orgaddress{\city{Fayetteville}, \postcode{72701}, \state{Arkansas}, \country{USA}}}

\affil[2]{\orgname{Wells Fargo}, \orgaddress{\city{New York}, \postcode{10001}, \state{New York}, \country{USA}}}

\affil[3]{\orgdiv{Department of Physics}, \orgname{University of Arkansas}, \orgaddress{\city{Fayetteville}, \postcode{72701}, \state{Arkansas}, \country{USA}}}

\affil[4]{\orgdiv{Department of Physics}, \orgname{Montana State University}, \orgaddress{\city{Bozeman}, \postcode{59717}, \state{Montana}, \country{USA}}}

\affil[5]{\orgdiv{Department of Electrical and Computer Engineering}, \orgname{Mississippi State University}, \orgaddress{\city{Starkville}, \postcode{39762}, \state{Mississippi}, \country{USA}}}

\affil[6]{\orgdiv{MonArk NSF Quantum Foundry}}


\abstract{
Parameterized Quantum Circuits (PQCs) have been acknowledged as a leading strategy to utilize near-term quantum advantages in multiple problems, including machine learning and combinatorial optimization. When applied to specific tasks, the parameters in the quantum circuits are trained to minimize the target function. Although there have been comprehensive studies to improve the performance of the PQCs on practical tasks, the errors caused by the quantum noise downgrade the performance when running on real quantum computers. In particular, when the quantum state is transformed through multiple quantum circuit layers, the effect of the quantum noise happens cumulatively and becomes closer to the maximally mixed state or complete noise. This paper studies the relationship between the quantum noise and the diffusion model. Then, we propose a novel diffusion-inspired learning approach to mitigate the quantum noise in the PQCs and reduce the error for specific tasks. Through our experiments, we illustrate the efficiency of the learning strategy and achieve state-of-the-art performance on classification tasks in the quantum noise scenarios.
}




\maketitle

\section{Introduction}\label{sec1}

Quantum machine learning (QML) \cite{biamonte2017quantum,ciliberto2018quantum,lloyd2013quantum,schuld2015introduction} is an emerging and promising interdisciplinary research direction in the fields of quantum computing and artificial intelligence. In this area, quantum computers are expected to enhance machine learning algorithms through their inherent parallel characteristics, thus demonstrating quantum advantages to solve some computational tasks out of reach even of classical supercomputers \cite{harrow2017quantum}. 
With the increasing enormous efforts from academia and industry, current quantum devices (usually acknowledged as the noisy intermediate-scale quantum (NISQ) devices \cite{preskill2018quantum}) already can show quantum advantages on specific carefully designed tasks \cite{arute2019quantum,zhong2020quantum} despite their limitations in quantum circuit width and depth. 
Moreover, prior experiments represent evidence for the utility of quantum computing on NISQ devices \cite{kim2023evidence}.
Thus, the NISQ devices open a direction to explore the quantum advantages of quantum machine learning tasks and leading strategies of hybrid classical-quantum algorithms, including parameterized quantum circuits.
Parameterized Quantum Circuits (PQCs) contain trainable parameters, offer a concrete way to implement algorithms, and demonstrate quantum supremacy in the NISQ era.
Even at low circuit depth, some classes of PQCs are capable of generating highly non-trivial outputs \cite{harrow2017quantum,benedetti2019parameterized}. 
For example, classical resources cannot efficiently simulate the class of PQCs called instantaneous quantum polynomial time under well-believed complexity-theoretic assumptions.
However, the behavior and impact of quantum noise remain critical questions for quantum computers today.
The NISQ devices still suffer from a high error rate of $10^{-2}$ to $10^{-4}$, much higher than CPUs/GPUs with an error rate of $10^{-18}$. 
The quantum errors, unfortunately, introduce a detrimental influence on PQCs accuracy.

Noise mitigation techniques \cite{temme2017error,wille2019mapping,li2017efficient,czarnik2021error,strikis2021learning} have been proposed to reduce the noise impact. 
However, these methods do not utilize the unique characteristics of PQCs and can only be applied to the inference process of the PQCs. 
Meanwhile, prior PQCs work \cite{farhi2018classification,benedetti2019parameterized,jiang2021co} does not study the impact caused by the quantum noise.

Diffusion models are inspired by non-equilibrium thermodynamics \cite{sohl2015deep,ho2020denoising,song2020denoising}. 
They are defined as a Markov chain of diffusion steps to gradually add random noise to data and then learn to reverse the diffusion process to construct desired data from the noise.
In quantum computing, random noise can be added into a quantum state by depolarizing until obtaining a maximally mixed state \cite{king2003capacity}.
Inspired by the quantum depolarizing channel, some prior works propose diffusion models on quantum computing \cite{cacioppo2023quantum,chen2024quantum,parigi2024quantum}.
However, more literature is needed to study the relationship between diffusion models and learnable quantum noise mitigation.

\noindent
\textbf{Contributions of this Work:}
In this work, we present the insights of diffusion-inspired modeling in the problem of PQCs. We show the relationship between diffusion and denoising processes in diffusion models and the forward and reverse processes in quantum states. The contributions of this paper are three-fold. First, we investigate the quantum noise properties in the PQCs and express the similarity between diffusion models and quantum computing in the noised PQCs. Then, we introduce a diffusion-inspired quantum noise mitigation framework for the PQCs. Second, from the diffusion-inspired quantum noise mitigation framework, we propose a novel loss function, i.e., forward-backward quantum divergence loss, to learn the quantum noise model for mitigation. Finally, our proposed method is benchmarked on various specific tasks and achieves State-of-the-Art (SOTA) results compared to the prior methods.

\section{Related Work}

\subsection{Quantum Machine Learning}
There has been recent interest in studying the combination between quantum computing and machine learning.
Early studies explored the quantum algorithms in linear machine learning, including clustering \cite{lloyd2013quantum,nguyen2023quantum,nguyen2024qclusformer,nguyen2024quantum}, principal component analysis \cite{lloyd2014quantum}, least-squares fitting \cite{schuld2016prediction,kerenidis2020quantum}, and classification \cite{rebentrost2014quantum,nguyen2024hierarchical}, to utilize the quantum speedup over classical machine learning algorithms.
Prior work focused on the quantum neural networks using the framework of variational quantum algorithms or parameterized quantum circuits \cite{panella2011neural,mitarai2018quantum}.
Cong et al. \cite{cong2019quantum} introduced the quantum convolutional neural network that extends the fundamental properties of classical CNNs to quantum computing while requiring less trainable parameters.
Meanwhile, Bausch \cite{bausch2020recurrent} proposed the quantum recurrent neural networks by utilizing the structure of the variational quantum eigensolver circuits.
Huang et al. \cite{huang2021experimental} presented hybrid quantum generative adversarial networks to generate data via quantum computers effectively.
Romero et al. \cite{romero2017quantum} introduced quantum autoencoders to reduce the dimensionality of quantum states.
Additionally, the quantum computing theory is applied to classical deep learning problems \cite{tang2022image,zhang2023quantum,nguyen2024quantumbrain}

\subsection{Quantum Noise Mitigation}
Prior studies on quantum noise in the quantum circuit and approaches to mitigate the errors caused by quantum noise exist.
Li and Benjamin \cite{li2017efficient} and Temme et al. \cite{temme2017error} introduced zero-noise extrapolation that tries to obtain zero-noise value by using data points at different circuit fault rates and computing the expected value at circuit fault rate zero.
Temme et al. \cite{temme2017error} also first presented the probabilistic error cancellation that reformulates the noise model as a linear combination and estimates this noise model via probability fitting with sampled quantum states.
Czarnik et al. \cite{czarnik2021error} and Strikis et al. \cite{strikis2021learning} applied learning-based methods to obtain the error-mitigated expectation value using training circuits.
QuantumNAT \cite{wang2022quantumnat} was introduced to reduce the error in the PQCs via post-measurement processing to mitigate the difference between quantum feature distributions in noise-free and noisy cases. However, the mitigation processing on quantum circuit operations is not considered.

\section{Background}

\subsection{Quantum Basics}
In general, quantum information is described by quantum states \cite{nielsen2001quantum}. 
An $n$-qubit quantum state is mathematically represented by a density matrix $\rho\in \mathbb{C}^{2^n\times 2^n}$ with property $\Tr(\rho)=1$. 
If $\operatorname{Rank}(\rho)=1$, the quantum state $\rho$ is a pure state; otherwise, it is a mixed state. 
A pure state can also be represented by a unit vector $\ket{\psi} \in \mathbb{C}^{2^n}$, where $\rho=\op{\psi}{\psi}$ and $\bra{\psi}=\ket{\psi}^\dagger$. 
A mixed state can be defined as a weighted sum of pure states and presented as in Eqn. \eqref{eqn:eq1}.
\begin{equation} \label{eqn:eq1}
    \rho = \sum_i \lambda_i\op{\psi_i}{\psi_i}, \quad
    \lambda_i \geq 0, \sum_i \lambda_i=1
\end{equation}
Specifically, a mixed state whose density matrix is proportional to the identity matrix is called the maximally mixed state $\mathbbm{1}_n = \frac{I}{2^n}$. Physically, it is a uniform mixture of states on an orthonormal basis. It means that all states occur with the same probability.

A quantum state $\rho$ can be evolved to another state $\rho^\prime$ through a quantum circuit (or gate) mathematically represented by a unitary matrix $U$, i.e., $\rho^\prime = U\rho U^\dagger$.
Typical single-qubit gates include Pauli gates:
\begin{equation}
    \sigma_x = \begin{bmatrix} 0 & 1\\ 1 & 0\end{bmatrix}, \quad
    \sigma_y = \begin{bmatrix} 0 & -i\\ i & 0\end{bmatrix}, \quad
    \sigma_z = \begin{bmatrix} 1 & 0\\ 0 & -1\end{bmatrix}
\end{equation}
and their corresponding rotation gates $R_{\sigma}(\theta) = e^{-i \theta \sigma/2}$ with a parameter $\theta$ and $\sigma \in \{\sigma_x, \sigma_y, \sigma_z\}$.
A multi-qubit gate can be either an individual gate (e.g., CNOT) or a tensor product of single-qubit gates.
To get classical information from a quantum state $\rho^\prime$, one needs to perform quantum measurements, e.g., calculating the expectation value $\langle H\rangle=\Tr(H\rho^\prime)$ of a Hermitian matrix $H$, and we often call $H$ an observable.

\subsection{Parameterized Quantum Circuits}
The parameterized quantum circuits (PQCs), also known as variational quantum circuits (VQCs) \cite{benedetti2019parameterized, mitarai2018quantum}, are a special kind of quantum circuit with parameters that can be optimized or learned iteratively.
The PQCs are composed of three parts, including data encoding, parameterized layer, and quantum measurements. 

PQCs use a hybrid quantum-classical procedure to optimize the trainable parameters iteratively.
The popular optimization approaches include gradient descent \cite{sweke2020stochastic}, parameter-shift rule \cite{wierichs2022general,mitarai2018quantum}, and gradient-free techniques \cite{nannicini2019performance,chen2022variational}.
All learning methods take the training data as input and evaluate the model performance by comparing the predicted and ground-truth labels. 
Based on this evaluation, the methods update the model parameters for the next iteration and repeat the process until the model converges and achieves the desired performance. 
The hybrid method performs the evaluation and parameter optimization on a classical computer, while the model inference is processed on a quantum computer.

\begin{figure}[t]
    \centering
    \includegraphics[width=\linewidth]{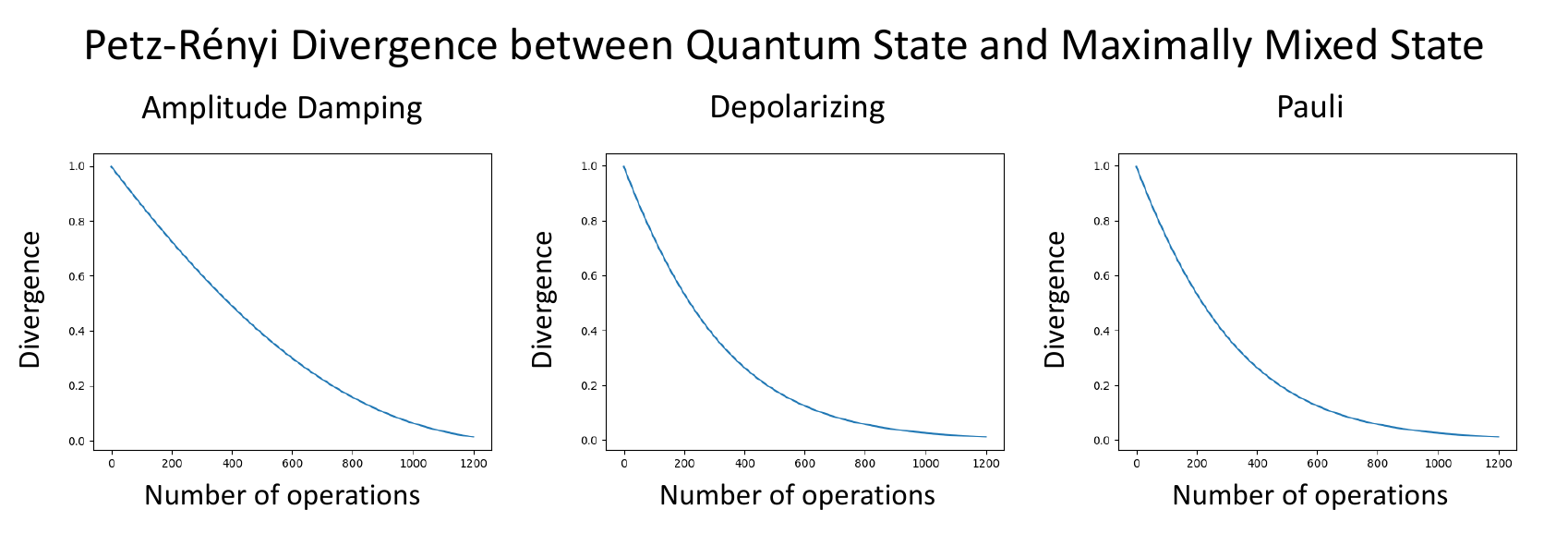}
    \caption{
    \textbf{The divergence between the quantum state and maximally mixed state when running operations on the IBM Quito quantum system.}
    After a number of quantum state operations with noise, the quantum state becomes closer to the full quantum noise, i.e., maximally mixed state. 
    It shows the relation between the effect of quantum noise and the number of quantum operations in PQCs.
    Note that although the amplitude damping makes the quantum state close to the $|0\rangle$ state, the quantum state distribution still gets closer to the maximally mixed state in the first 1200 operations.}
    \label{fig:pr_div}
\end{figure}

\subsection{Noise in Quantum Computing}
Noise refers to the multiple factors that can affect the accuracy of the calculations a quantum computer performs. 
Because of the noise in quantum computing, the transformation of the quantum state can cause errors.
Pauli channel error is one of the common noise models that cause probabilistic error between Pauli operations, defined as
$\tilde{\rho} = \tilde{\Lambda}(\rho) = \rho + \sum_{\sigma \in \mathcal{K}} \lambda_\sigma(\sigma \rho \sigma^\dagger - \rho)$, 
where $\mathcal{K}$ is a set of Pauli operations and $\lambda_\sigma$ is the probability that the error caused by $\sigma$.
Here, we define $\rho_i$ as a noise-free quantum state and $\tilde{\rho}_i$ as a quantum state in the noise scenario.
These errors are cumulated via quantum circuit layers that make the quantum state closer to the maximally mixed state, i.e., the fully noise quantum state as shown in Fig. \ref{fig:pr_div}.
Moreover, the noise model $\tilde{\Lambda}(\rho)$ can be formulated to an alternative model $\Lambda(\rho)$:
\begin{equation}
    \Lambda(\rho) = 
    \prod_{\sigma \in \mathcal{K}}(w_\sigma \cdot + (1 - w_\sigma)\sigma \cdot \sigma^\dagger) \rho
\label{eq:noise_linear}
\end{equation}
where $w_\sigma = 2^{-1} (1 + e^{-2\lambda_\sigma})$.
In the PQCs, the error caused by the noise can occur in the data encoding, parameterized layer, or even the quantum measurements.
However, as the quantum operations are primarily in the parameterized layers, we focus on the quantum noise mitigation in this part of the PQCs.

\section{Diffusion-inspired Modeling for Quantum Denoising}

\subsection{Diffusion Modeling Revisited}

Diffusion models \cite{ho2020denoising} are a class of latent variable models that learn a generative model to reverse a fixed probabilistic noising process $\mathbf{x}_0 \rightarrow \mathbf{x}_1 \rightarrow \dots \rightarrow \mathbf{x}_T$, where $\mathbf{x}_1, \dots, \mathbf{x}_T$ are latent variables of the same dimensionality of the data $\mathbf{x}_0 \sim q(\mathbf{x}_0)$.
This probabilistic noising process gradually adds noise to clean data $\mathbf{x}_0$ until no information remains, i.e., pure noise $\mathbf{x}_T \sim p(\mathbf{x}_T)$. 
The divergence of data in this process can be formulated as:
\begin{equation}
    D(q(\mathbf{x}_0) || \mathcal{N}_G) > D(q(\mathbf{x}_1) || \mathcal{N}_G) > \dots > D(q(\mathbf{x}_T) || \mathcal{N}_G)
\label{eq:classical_gaussian_noise_div}
\end{equation}
where $\mathcal{N}_G$ is the probability distribution of the pure Gaussian noise.
In most cases, the Kullback-Leibler divergence is applied to compute the difference between distributions of classical data.
For continuous data, the forward process is defined as a fixed Markov chain $q(\mathbf{x}_t | \mathbf{x}_{t-1})$ with Gaussian transitions.
The Markov chain of the reverse process can be obtained by approximating the true posterior $q(\mathbf{x}_{t-1} | \mathbf{x}_0, \mathbf{x}_t)$ with a model $p_\theta(\mathbf{x}_{t-1} | \mathbf{x}_t)$.
Therefore, sampling new data $\mathbf{x}_0$ from the modeled data distribution $p_\theta(\mathbf{x}_0) = p(\mathbf{x}_T) \prod_{t=1}^T p_\theta(\mathbf{x}_{t-1} | \mathbf{x}_t)$ is performed by starting from random noise $\mathbf{x}_T \sim p(\mathbf{x}_T)$ and gradually denoising it over $T$ steps $\mathbf{x}_T \rightarrow \mathbf{x}_{T-1} \rightarrow \dots \rightarrow \mathbf{x}_0$.

\subsection{Diffusion-inspired Quantum Noise Mitigation}

\begin{figure}[t]
    \centering
    \includegraphics[width=0.95\linewidth]{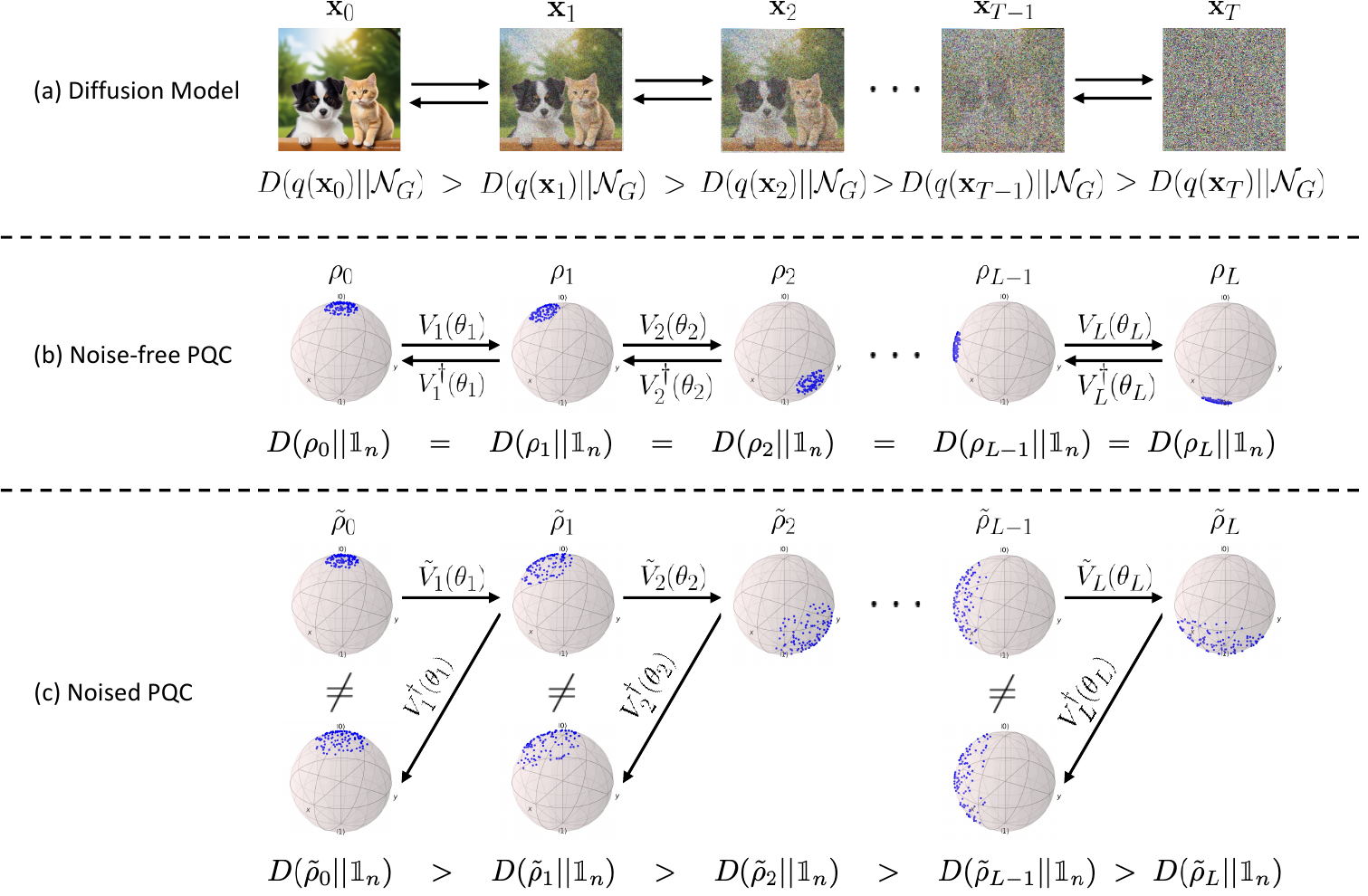}
    \caption{
    (a) In the traditional diffusion model, the distribution of the data $q(\mathbf{x}_i)$ becomes closer to the Gaussian noise $\mathcal{N}_G$ through diffusion steps.
    (b) In the noise-free scenario, each quantum transformation $V_i(\theta_i)$ can be reversed and maintain the quantum information of the previous circuit layer via reverse operation $V_i^\dagger(\theta_i)$. 
    (c) In the PQCs having quantum noise case, the quantum state transformations $\tilde{V}_i(\theta_i) = \Lambda_i \circ V_i(\theta_i)$ are affected by noise $\Lambda_i$ that makes the reversed quantum state different from the original one. Motivated by this, we propose a learning-based method to mitigate the quantum noise via computing the divergence between the quantum state and its noisy forward-and-backward state. 
    }
    \label{fig:quantum_state_reverse}
\end{figure}

Quantum computing has similar properties to diffusion models when transforming the quantum state through quantum circuits. In the following, we derive a diffusion-inspired learning for quantum noise mitigation.
The two main components of this learning strategy are forward and reverse processes. 
The forward process transforms a quantum state into a next state through a quantum circuit layer and adds noise to the quantum state as the nature of PQCs in the noise scenario. As the resulting quantum state has noise, a denoising process is considered to reduce the error. 
In the general diffusion models, this process is solved by the reverse process to train the denoising module.

Given a quantum state $\rho_{i-1}$ and a transformed quantum state $\rho_i$ from $\rho_{i-1}$, without noise, 
as the Shannon entropy of the quantum state is unchanged by the unitary circuit $S(\rho) = S(U \rho U^\dagger)$,
the divergence between $\rho_i$ and the maximally mixed state $\mathbbm{1}_n$ is the same as the divergence between $\rho_{i-1}$ and $\mathbbm{1}_n$: 
\begin{equation}
    D(\rho_i || \mathbbm{1}_n) = D(\rho_{i-1} || \mathbbm{1}_n)
\end{equation}
However, as shown in Fig. \ref{fig:pr_div}, in the noise case, the quantum noise makes the divergence of the noise quantum state $\tilde{\rho}_i$ closer to the maximally mixed state $\mathbbm{1}_n$ after the quantum transformation:
\begin{equation}
    D(\tilde{\rho}_i || \mathbbm{1}_n) < D(\tilde{\rho}_{i-1} || \mathbbm{1}_n)
\end{equation}
This phenomenon is similar to the forward process of the diffusion model when the classical data becomes closer to the Gaussian noise, as shown in Eqn. \eqref{eq:classical_gaussian_noise_div}.
Moreover, as shown in Eqn. \eqref{eq:noise_linear}, the quantum noise in the PQCs can be considered as a Markov chain analogous to the diffusion process.

On the other hand, as the quantum circuits are reversible, the quantum state $\rho_i$ can be backward into the quantum state $\rho_{i-1}$ of the previous quantum circuit layer in the noise-free scenario.
In the quantum noise scenario, as the quantum state $\tilde{\rho}_i$ has noise, the backward process causes errors that the resulting quantum state is not similar to its original state. 
Hence, a noise mitigation module models and reduces the quantum noise, making the backward quantum state closer to the original state.
Fig. \ref{fig:quantum_state_reverse} illustrates the forward and backward processes. From this observation, we proposed a diffusion-inspired learning method to model the quantum noise distribution for noise mitigation.

\section{The Proposed Method}

To address the quantum noise in the PQCs, we first consider the general PQCs framework and their quantum noise mitigation module in the noise scenario.
Then, we study the distribution of the quantum noise and quantum states. Finally, we introduce a novel Forward-backward Quantum Divergence Loss for quantum noise mitigation.

\begin{figure}
    \centering
    \includegraphics[width=0.9\linewidth]{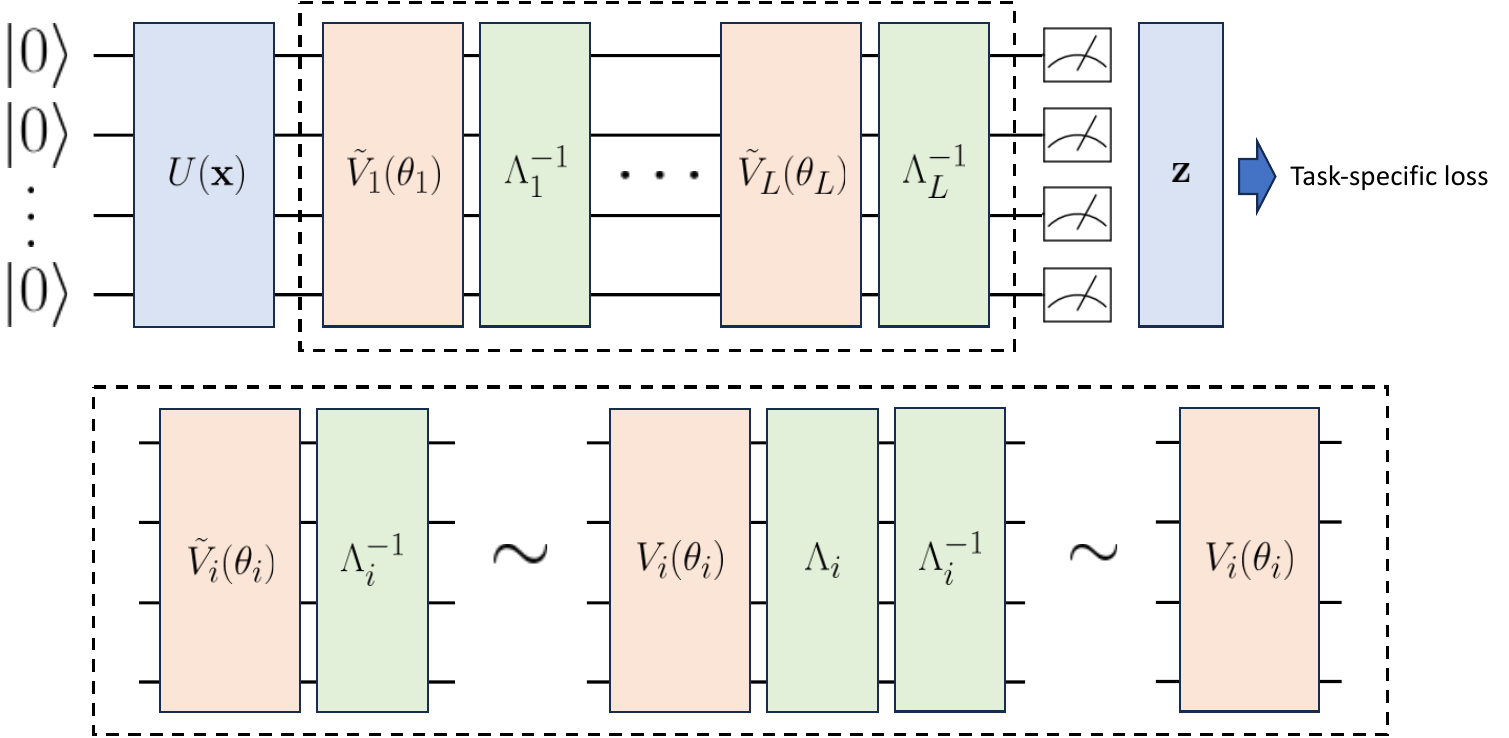}
    \caption{
    An overall framework of PQCs with quantum noise mitigation. 
    For every quantum circuit layer $\tilde{V}_i(\theta_i) = \Lambda_i \circ V_i(\theta_i)$, a quantum noise mitigation layer $\Lambda_i^{-1}$ is applied to reduce the error.
    }
    \label{fig:overall_framework}
\end{figure}

\subsection{The Overall Framework}

Fig. \ref{fig:overall_framework} illustrates the overall framework of the quantum noise mitigation in PQCs.
Given classical data $\mathbf{x}$, a data encoder $U$ is applied to encode the classical data into a quantum state $\rho_0$. Then, the PQCs are trained to transform and measure the quantum state for a specific task. In the noise-free scenario, learnable unitary matrices $V_i(\theta_i)$ are applied to transform the quantum state:
\begin{equation}
    \rho_i = V_i(\theta_i) \rho_{i-1} V_i^\dagger(\theta_i)
\end{equation}
The design of the parameterized circuits is described in Section \ref{sec:experimental_results}.
Because of the noise in quantum computing, we can model the noise by a matrix $\Lambda_i$:
\begin{equation}
    \tilde{\rho}_i = \tilde{V}_i(\theta_i) \tilde{\rho}_{i-1} \tilde{V}_i^\dagger(\theta_i), \quad
    \tilde{V}_i(\theta_i) = \Lambda_i \circ V_i(\theta_i)
\label{eq:noise_pqc}
\end{equation}
It causes errors in the PQCs when training and inferring. Hence, quantum noise mitigation layers $\Lambda_i^{-1}$ are applied to reduce these errors.
To obtain these quantum noise mitigation layers, the probabilities of the quantum noise have to be computed.
Therefore, in this work, we aim to learn the noise model $\Lambda_i$ to mitigate the error caused by the quantum noise. 

\subsection{Quantum Noise Distribution Learning}
\label{sec:distribution_learning}
As we want to learn the quantum noise model of $\Lambda_i$, we define a probability density function $p_\omega(\Lambda_i)$ where $\omega$ are learnable parameters of the overall model.
Moreover, to train the PQCs for the specific task, we also learn the distribution of the quantum states $\tilde{\rho}_i$ where $i \in \{0, 1, \dots L-1\}$ to represent the data for specific tasks.
Since $\Lambda_i$ and $\tilde{\rho}_{i-1}$ are independent, the objective of $\omega$ can be defined as:
\begin{equation}
\begin{split}
    \omega^* = \arg\min_\omega \sum_{i=1}^L \left( - \log p_\omega(\Lambda_i) - \log p_\omega(\tilde{\rho}_{i-1}) \right) 
    = \arg\min_\omega \sum_{i=1}^L - \log p_\omega(\Lambda_i, \tilde{\rho}_{i-1})
\end{split}
\label{eq:objective_noise}
\end{equation}
We also define the forward process from the quantum state $\tilde{\rho}_{i-1}$ to the next quantum states $\tilde{\rho}_i$ as $q(\tilde{\rho}_i | \tilde{\rho}_{i-1})$.
To optimize the objective function Eqn. \eqref{eq:objective_noise}, we optimize the upper bound on negative log-likelihood via Jensen's inequality:
\begin{equation}
\begin{split}
    \sum_{i=1}^L - \log p_\omega(\Lambda_i, \tilde{\rho}_{i-1})
    &\leq \mathbb{E}_q\left[
    - \sum_{i=1}^L \log \frac{p_\omega(\Lambda_i, \tilde{\rho}_{i-1}, \tilde{\rho}_i)}{q(\tilde{\rho}_i | \tilde{\rho}_{i-1})}
    \right] := \mathcal{L}
\label{eq:jensen_inequality}
\end{split}
\end{equation}
Then $\mathcal{L}$ can be expanded as:
\begin{equation}
\begin{split}
    \mathcal{L}
    &= \mathbb{E}_q\left[
    \underbrace{
    \sum_{i=1}^L D_{KL}\left(
    p_\omega(\tilde{\rho}_{i-1} | \Lambda_i, \tilde{\rho}_i) || q(\tilde{\rho}_{i-1} | \tilde{\rho}_i)
    \right)
    }_{\mathcal{L}_{fb}}
    \underbrace{
    - \sum_{i=1}^L \log p_\omega(\Lambda_i, \tilde{\rho}_i)
    }_{\mathcal{L}_{task}}
    - \log \frac{q(\tilde{\rho}_{0})}{q(\tilde{\rho}_L)}
    \right]
\end{split}
\label{eq:kl_expand}
\end{equation}
(See Appendix for details of Eqn. \eqref{eq:jensen_inequality} and Eqn. \eqref{eq:kl_expand}.) The Kullback-Leibler divergence in Eqn. \eqref{eq:kl_expand} compares the original quantum state and its forward-backward in the quantum circuit layer. Thus, we propose a learning method to model the quantum noise via quantum divergence.

\subsection{Forward-backward Quantum Divergence Loss}

\begin{figure}[t]
    \centering
    \includegraphics[width=0.7\linewidth]{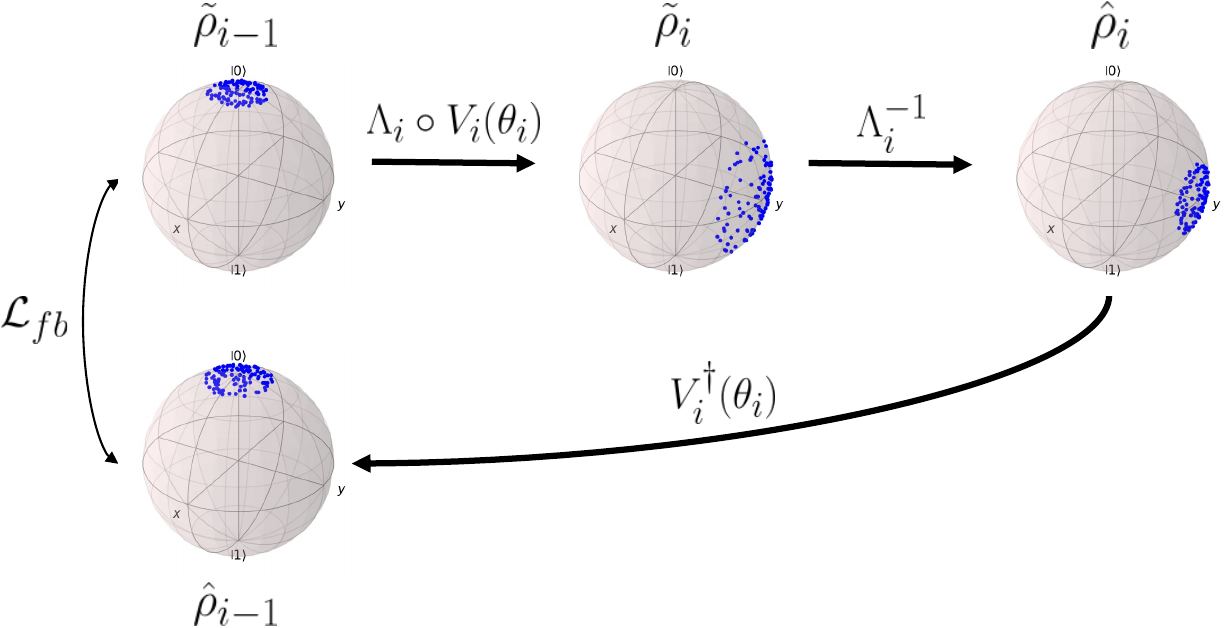}
    \caption{
    \textbf{The forward-backward quantum divergence loss in a quantum circuit.}
    For each layer $V_i(\theta_i)$ of the quantum circuit, the current quantum state $\tilde{\rho}_i$ is denoised by a learnable noise mitigation layer $\Lambda_i^{-1}$. 
    }
    \label{fig:fb_loss}
\end{figure}

In quantum information theory, to measure the similarity between two quantum states, fidelity is defined as the probability that one state will pass a test to identify as the other. In this work, we utilize the fidelity of quantum states to define a forward-backward quantum divergence loss.
Let $F(\rho, \sigma) \in [0, 1]$ be the fidelity between quantum states $\rho$ and $\sigma$, then $F(\rho, \sigma) = 1$ when $\rho$ and $\sigma$ are completely similar and vice versa.

Given a quantum state $\tilde{\rho}_{i-1}$ and its forwarded quantum state $\tilde{\rho}_i$ as Eqn. \eqref{eq:noise_pqc}, we define a learnable noise mitigation layer $\Lambda^{-1}_i(\cdot, \omega_i)$ to reduce the noise from quantum state $\tilde{\rho}_i$.
\begin{equation}
    \hat{\rho}_i = \Lambda^{-1}_i(\tilde{\rho}_i, \omega_i)
\end{equation}
As the quantum circuits are reversible, the quantum state $\hat{\rho}_i$ can be reversed to the previous state via the quantum circuit $V_i^\dagger(\theta_i)$. 
\begin{equation}
    \hat{\rho}_{i-1} = V_i^\dagger(\theta_i) \hat{\rho}_i V_i(\theta_i)
\end{equation}
Note that in practice, the inverse circuits also have noises.
However, each quantum operation on each qubit causes a specific noise, and its inverse operation also causes a similar noise.
Because of that, the quantum noise learning of the inverse circuits is similar to the forward circuits.
Hence, for simplification, we assume that the inverse circuits are noise-free.

Finally, we define a forward-backward quantum divergence loss to compute the similarity between quantum states $\tilde{\rho}_{i-1}$ and $\hat{\rho}_{i-1}$ as:
\begin{equation}
    \mathcal{L}_{fb}(\tilde{\rho}_{i-1}, \hat{\rho}_{i-1}) = 
    - \log F(\tilde{\rho}_{i-1}, \hat{\rho}_{i-1})
\end{equation}
where $F(\rho, \sigma)$ is the fidelity between two quantum states $\rho$ and $\sigma$.
As the forward-backward procedure can be processed in $L$ quantum circuit layers, the overall forward-backward quantum divergence loss is computed as:
\begin{equation}
    \mathcal{L}_{fb} = \frac{1}{L} \sum_{i=1}^L \mathcal{L}_{fb}(\tilde{\rho}_{i-1}, \hat{\rho}_{i-1}) = 
    - \frac{1}{L} \sum_{i=1}^L \log F(\tilde{\rho}_{i-1}, \hat{\rho}_{i-1})
\end{equation}
The forward-backward quantum divergence loss process is illustrated in Fig. \ref{fig:fb_loss}.
To compute the fidelity between quantum states, a widely used metric is the quantum Rényi divergence \cite{muller2013quantum}:
\begin{equation}
    F(\rho, \sigma) = D(\rho || \sigma) = 
    2 \log \textrm{Tr}\left[\sqrt{\sqrt{\sigma}\rho\sqrt{\sigma}}\right]
\end{equation}

\subsection{Task-specific Training}

In addition to learning the quantum noise model, the PQCs are also trained for specific tasks, such as classification and clustering.
As shown in Eqn. \eqref{eq:kl_expand}, the PQCs can be trained to obtain a desired distribution of quantum states while learning the quantum noise model.
Thus, a task-specific loss $\mathcal{L}_{task}$ is defined for the PQCs.
Let $H$ be the observable of the PQC, a classical information obtained from the PQCs is measured as $\langle H \rangle = \textrm{Tr}(H \hat{\rho}_L)$.
Given $n$ qubits for the PQC, we define $n$ observables $H_i = I^{\otimes i} \otimes \sigma_z \otimes I^{\otimes (n - i - 1)}$ where $I \in \mathbb{R}^{2 \times 2}$ is the identity matrix and $\sigma_z$ is the Pauli-Z matrix.
A classical information vector $\mathbf{z} = \{\langle H_i \rangle\}_{i=0}^{n-1}$ is defined for inference and task-specific loss computing.
In this work, the PQCs are applied for the classification tasks.
Then the task-specific loss $\mathcal{L}_{task}$ is defined as:
\begin{equation}
    \mathcal{L}_{task} = - \sum_{i=0}^c \hat{y}_i \log y_i
\end{equation}
where $c$ is the number of classes, $\hat{y}_i \in \{0, 1\}$ is the classification ground-truth, and $\mathbf{y} = f(\mathbf{z})$ is the prediction from classical information vector $\mathbf{z}$.
Finally, the total loss of the quantum noise mitigation training for the PQCs is computed as:
\begin{equation}
    \mathcal{L}_{total} = \alpha_{fb} \mathcal{L}_{fb} + \alpha_{task} \mathcal{L}_{task}
\end{equation}
where $\alpha_{fb}, \alpha_{task} \in \mathbb{R}$ is the training hyperparamter.

\section{Experimental Results}
\label{sec:experimental_results}

In this section, we evaluate the proposed method in the quantum noise context.
We first describe the experiment setups, including datasets, implementation, and evaluation protocol.
Then, we present the ablation studies to illustrate the effectiveness of our proposed method.
Finally, we illustrate the numerical results of the proposed method and compare our approach with prior quantum noise mitigation methods.

\subsection{Experiment Setups}
\label{sec:experiment_setups}

\begin{wraptable}[15]{r}{0.5\textwidth}
\centering
\caption{
    \textbf{Effectiveness of our approach on the MNIST-4 benchmark.}
    We compute the mean accuracies (\%) and their standard deviation to evaluate the approach with different circuit designs, i.e., RX + CNOT, U2 + CNOT, and U3 + CNOT, with different noise mitigation step sizes, and without or with the forward-backward quantum divergence loss $\mathcal{L}_{fb}$.
}
\setlength{\tabcolsep}{5pt}
\resizebox{\linewidth}{!}{
\begin{tabular}{c|c|ccc}
\Xhline{2\arrayrulewidth}
\multirow{2}{*}{\textbf{Circuits}} & \multirow{2}{*}{$\mathcal{L}_{fb}$} & \multicolumn{3}{c}{\textbf{Step size}} \\
 & & 4 & 2 & 1 \\
\hline
\multirow{2}{*}{RX + CNOT}
&                  & $41.76 \pm 1.46$ & $41.90 \pm 1.53$ & $42.61 \pm 1.37$ \\
& \checkmark       & $44.30 \pm 1.65$ & $45.55 \pm 1.46$ & $45.89 \pm 1.43$ \\
\hline
\multirow{2}{*}{U2 + CNOT}
&                  & $40.58 \pm 1.63$ & $42.91 \pm 1.78$ & $43.17 \pm 1.57$ \\
& \checkmark       & $44.42 \pm 1.69$ & $45.50 \pm 1.73$ & $46.44 \pm 1.60$ \\
\hline
\multirow{2}{*}{U3 + CNOT}
&                  & $42.51 \pm 1.46$ & $43.18 \pm 1.47$ & $43.37 \pm 1.39$ \\
& \checkmark       & $45.36 \pm 1.29$ & $46.20 \pm 1.45$ & $46.37 \pm 1.20$ \\
\Xhline{2\arrayrulewidth}
\end{tabular}
}
\label{tab:abl_studies_step}
\end{wraptable}

\textbf{Datasets. }
Following \cite{wang2022quantumnat}, we evaluate the proposed method on four classification tasks, including MNIST \cite{lecun2010mnist} 4-class (0, 1, 2, 3) and 2-class (3, 6); and Fashion \cite{xiao2017fashion} 4-class (t-shirt/top, trouser, pullover, dress) and 2-class (dress, shirt).
The images are resized into $8 \times 8$ and encoded via phase encoding using multiple rotation circuits.

\noindent
\textbf{Implementation. }
This work uses the quantum simulation for the PQCs and the quantum noise.
For each layer, different learnable circuits are applied, including RX, U2 (RX + RY), and U3 (RX + RY + RZ).
The efficiency of each circuit is shown in the ablation studies.
Then, a non-learnable circuit, i.e., controlled-NOT (CNOT), is used.
We use four qubits for the experiment and run simulations of the IBM quantum systems via Qiskit SDK \cite{Qiskit} on a Quadro RTX 8000 GPU.
In our experiments, we set $\alpha_{\text{fb}} = \alpha_{\text{task}} = 1$ and apply Adam \cite{kingma2014adam} as an optimizer with a learning rate of $2 \times 10^{-3}$.
The PQCs model training and testing is implemented based on the TorchQuantum library \cite{wang2022torchquantum}.

\begin{wrapfigure}[16]{r}{0.5\linewidth}
\vspace{-4mm}
    \centering
    \includegraphics[width=0.95\linewidth]{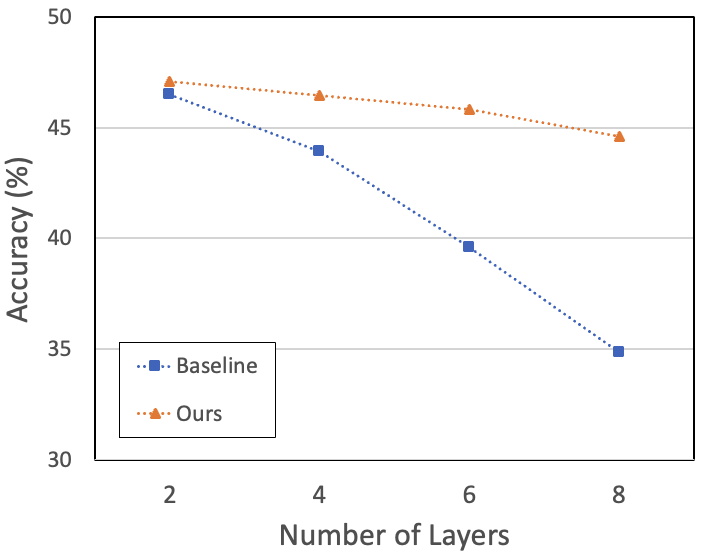}
    \caption{Ablation studies on different numbers of circuit layers.}
    \label{fig:abl_num_layers}
\end{wrapfigure}

\noindent
\textbf{Evaluation Protocol. }
For evaluation, we compare the proposed method with other settings, i.e., training and testing on noise-free PQCs, training on noise-free PQCs and testing on noised PQCs, and training and testing on noised PQCs as a baseline.
We also reimplement the training and evaluation of the prior quantum noise mitigation methods, i.e., QuantumNAT \cite{wang2022quantumnat} normalizing and quantizing the measurement to match the output distribution with the noise-free case, and Van Den Berg et al. \cite{van2023probabilistic} learning the probabilistic quantum noise model by sampling random quantum states.
The average accuracy metric is used in our experiments.
The experiments are processed five times for each setting, and the standard deviation is used to compute the variance of the results.

\subsection{Ablation Studies}
\label{sec:ablation_studies}

Our ablative experiments study the effectiveness of our proposed method on the performance of the PQCs on the MNIST-4 benchmark.

\noindent
\textbf{Effectiveness of the Forward-backward Quantum Divergence Loss.}
We evaluate the impact of Forward-backward Quantum Divergence Loss ($\mathcal{L}_{fb}$) in improving the performance of the PQCs in the noise scenario.
To demonstrate the efficiency of our proposed approach with diverse quantum circuit designs, we evaluate it with three different circuit designs: RX + CNOT, U2 + CNOT, and U3 + CNOT.
As shown in Table \ref{tab:abl_studies_step}, the forward-backward loss $\mathcal{L}_{fb}$ has significant improvements compared to using the noise mitigation with the task-specific loss $\mathcal{L}_{task}$ only.
Note that we can optimize the quantum noise mitigation layer with $\mathcal{L}_{\text{task}}$ only.
With the task loss, the noise mitigation module is optimized for the specific task without noise mitigation by itself.
In particular, the accuracy of the three circuit designs, i.e., RX + CNOT, U2 + CNOT, and U3 + CNOT, has been improved from $42.61\%$ to $45.89\%$, from $43.17\%$ to $46.44\%$, and from $43.37\%$ to $46.37\%$, respectively.
It shows that the forward-backward loss $\mathcal{L}_{fb}$ helps to learn the noise distribution better for quantum noise mitigation.

\noindent
\textbf{Effectiveness on Different Noise Mitigation Step Sizes.}
We investigate the impact of the quantum noise mitigation step size on the performance.
To achieve this, we conduct experiments on the MNIST-4 benchmark with a 4-layer PQCs model using three different step sizes, i.e., 4, 2, and 1.
In this case, step size 4 means we forward the quantum state $\tilde{\rho}_0$ through 4 layers of the PQCs and backward the resulting state $\tilde{\rho}_4$ to the initial state $\hat{\rho}_0$ for the forward-backward quantum divergence loss $\mathcal{L}_{fb}$.
Similarly, step size 2 means we forward the quantum state $\tilde{\rho}_i$ through 2 layers and backward the resulting state $\tilde{\rho}_{i+2}$ for the forward-backward loss $\mathcal{L}_{fb}$, and step size 1 means we compute the loss for every layer.
As depicted in Table \ref{tab:abl_studies_step}, the performance gradually increases when the step size is decreased for all different circuit designs.
It shows that quantum noise mitigation works effectively when the number of operation circuits for each process is negligible.

\noindent
\textbf{Effectiveness on Different Numbers of Circuit Layers.}
We evaluate the robustness of the proposed quantum noise mitigation approach on different numbers of circuit layers of the PQCs.
As illustrated in Fig. \ref{fig:abl_num_layers}, when the number of layers is increased, the performance of the baseline is dramatically dropped. Meanwhile, our proposed approach shows a slight decline in accuracy.
Hence, the proposed approach can mitigate the quantum noise effectively even if the PQCs have many operation circuits.

\begin{table}[t]
\centering
\caption{
\textbf{Experimental accuracies (\%) on 2- and 4-class benchmarks.} 
A noise-free testing is defined as training and testing on a noise-free PQC while the noise-free trained model is tested on the quantum noise scenario. 
We show the mean accuracies and their standard deviation for variances of the results.
}
\begin{tabular}{l|cccc}
\Xhline{2\arrayrulewidth}
\textbf{Method} & \textbf{MNIST-4} & \textbf{MNIST-2} & \textbf{Fashion-4} & \textbf{Fashion-2} \\
\hline
Noise-free testing                               & $49.87 \pm 0.31$ & $92.99 \pm 0.19$ & $49.53 \pm 0.35$ & $74.60 \pm 0.47$ \\
Noise-free training                              & $37.24 \pm 1.88$ & $79.07 \pm 8.22$ & $39.08 \pm 1.88$ & $66.35 \pm 3.39$ \\
\hline
Baseline                                         & $43.06 \pm 1.06$ & $82.37 \pm 4.62$ & $43.70 \pm 1.43$ & $68.50 \pm 3.12$ \\
QuantumNAT \cite{wang2022quantumnat}             & $43.59 \pm 1.15$ & $82.32 \pm 4.68$ & $43.60 \pm 1.40$ & $68.55 \pm 3.11$ \\
Van Den Berg et al. \cite{van2023probabilistic}  & $45.15 \pm 1.45$ & $84.32 \pm 4.30$ & $45.25 \pm 1.89$ & $69.15 \pm 2.44$ \\
\hline
\textbf{Ours}                                    & $\mathbf{46.44 \pm 1.60}$ & $\mathbf{85.32 \pm 4.35}$ & $\mathbf{46.83 \pm 1.81}$ & $\mathbf{70.50 \pm 2.57}$ \\
\Xhline{2\arrayrulewidth}
\end{tabular}
\label{tab:exp_results}
\end{table}

\subsection{Evaluation Results}
\label{sec:evaluation_results}

As shown in Table \ref{tab:exp_results}, our method outperforms previous methods evaluated on the two datasets, i.e., MNIST and Fashion.
In particular, in the MNIST dataset, our method achieves the accuracy of $46.44\%$ and $85.32\%$ on the MNIST-4 and MNIST-2 benchmarks, respectively, which shows better than the previous methods.
Meanwhile, the results for the Fashion dataset are $46.83\%$ and $70.50\%$ for the Fashion-4 and Fashion-2 benchmarks.

\section{Conclusions}

This paper has studied the quantum noise in PQCs and introduces a novel, learnable quantum noise mitigation approach to improve their robustness when running on NISQ devices.
By revisiting the diffusion model and investigating the relationship between the diffusion and denoising processes and the forward and reverse processes in quantum states, we have shown the quantum noise distribution learning ability while training the PQCs for specific tasks.
Hence, a novel forward-backward quantum divergence loss function has been introduced to learn the quantum noise model for quantum noise mitigation.
The experimental results on various benchmarks have shown our state-of-the-art method.

\section{Discussion}

\noindent
\textbf{Limitations:}
Our paper has chosen specific configurations of quantum systems and hyperparameters to support our hypothesis theoretically in the simulation.
However, other aspects, such as non-Markovian noise models, computational complexity, or learning hyperparameters, have yet to be thoroughly investigated.

\noindent
\textbf{Broader Impact:}
This work studies a diffusion-inspired approach to quantum noise model learning for noise mitigation in the PQCs.
Our contributions emphasize the importance of quantum noise mitigation in quantum machine learning and provide a solution to reduce the error and increase the robustness of the PQCs when running in real quantum computers.

\section*{Declarations}


\begin{itemize}
\item Funding: There is no funding for the manuscript.
\item Data availability: 
The MNIST dataset \cite{lecun2010mnist} is public available at \href{http://yann.lecun.com/exdb/mnist}{http://yann.lecun.com/exdb/mnist}.
The Fashion dataset \cite{xiao2017fashion} is public available at \href{https://github.com/zalandoresearch/fashion-mnist}{https://github.com/zalandoresearch/fashion-mnist}.
\item Code availability: The code will be available upon acceptance.
\item Author contribution: 
H.N wrote the main manuscript and prepared pseudo code, result tables, and experiment setups. 
S.C, H.C, N.B, and S.K provided fundamental materials for quantum information and computing. 
X.N and K.L discussed the novelty and the research direction. 
All the authors revised the manuscript. 
\end{itemize}

\newpage

\begin{appendices}

\section{Quantum Noise Mitigation Module}

In this work, we take account of the Pauli-Lindblad noise model for quantum noise mitigation.
As shown in Eqn. \eqref{eq:noise_linear}, for each Pauli channel, the noise model is defined as $(w_\sigma \rho + (1 - w_\sigma) \sigma \rho \sigma)$ where $w_\sigma = 2^{-1}(1 + e^{-2\lambda_\sigma})$.
Then, the inverse of the Pauli channel can be written as $(2 w_\sigma - 1)^{-1} (w_\sigma \tilde{\rho} - (1 - w_\sigma) \sigma \tilde{\rho} \sigma)$.
In overall, the quantum noise $\Lambda(\cdot)$ can be mitigated from the quantum noise mitigation layer $\Lambda^{-1}(\cdot)$ as:
\begin{equation}
    \Lambda^{-1}(\tilde{\rho}) = \gamma \prod_{\sigma \in \mathcal{K}}(w_\sigma \cdot - (1 - w_\sigma) \sigma \cdot \sigma^\dagger) \tilde{\rho}
\end{equation}
where $\gamma = \prod_{\sigma \in \mathcal{K}} (2 w_\sigma - 1)^{-1} = \exp(\sum_{\sigma \in \mathcal{K}} 2 \lambda_\sigma)$ is \textit{sampling overhead}.
In this case, $\{\lambda_\sigma\}_{\sigma \in \mathcal{K}}$ are learnable parameters for the quantum noise mitigation.

\section{Quantum Noise Mitigation Learning Algorithm}

The quantum noise mitigation learning process can be described in Algorithm \ref{algo:pseudo_code}.

\RestyleAlgo{ruled}
\begin{algorithm}[h]
\caption{Pseudo-code for the implementation of Quantum Noise Mitigation}
\SetKwComment{Comment}{ // }{}

\KwData{
\\$\{\mathbf{x}_i\}_{i=1}^{N}$ : a set of $N$ classical inputs
\\$\{\hat{y}_i\}_{i=1}^{N} \in \mathbb{R}^N$ : a set of $N$ labels
\\$n$ : the number of qubits
\\$L$ : the number of layers in the PQCs
\\$\alpha_{fb}$, $\alpha_{task}$ : training hyperparameters
}

\While{not convergent} {

    $\tilde{\rho}_0 \gets U(\mathbf{x})$
    \Comment{Encode the classical data into a quantum state}

    $\mathcal{L}_{fb} \gets 0$
    \Comment{Initialize the forward-backward quantum divergence loss to zero}

    \For{$i \in [1..L]$} {
        $\tilde{\rho}_i \gets \tilde{V}_i(\theta_i) \tilde{\rho}_{i-1} \tilde{V}_i^\dagger(\theta_i)$ 
        \Comment{Forward the quantum state in the noised case}

        $\hat{\rho}_i \gets \Lambda_i^{-1}(\tilde{\rho}_i)$ 
        \Comment{Reduce the quantum noise}

        $\hat{\rho}_{i-1} \gets V_i^\dagger(\theta_i) \hat{\rho}_i V_i(\theta_i)$ 
        \Comment{Backward the quantum state}

        $\mathcal{L}_{fb} \gets \log F(\tilde{\rho}_{i-1}, \hat{\rho}_{i-1})$ 
        \Comment{Compute the forward-backward loss}
    }
    
    $\mathcal{L}_{fb} \gets \frac{1}{L}\mathcal{L}_{fb}$ 
    \Comment{Compute the average of the loss}

    $\mathbf{z} \gets \{\langle H_i \rangle\}_{i=0}^{n-1}$
    \Comment{Compute the classical information from the quantum state $\hat{\rho}_L$}

    $\mathbf{y} \gets f(\mathbf{z})$
    \Comment{Compute the task-specific output}

    $\mathcal{L}_{task} \gets - \sum_{i=0}^c \hat{y}_i \log y_i$
    \Comment{Compute the task-specific loss}

    $\mathcal{L}_{total} \gets \alpha_{fb} \mathcal{L}_{fb} + \alpha_{task} \mathcal{L}_{task}$
    \Comment{Compute the total loss}
    
    $\theta \gets \theta - \lambda \nabla_{\theta}\mathcal{L}_{total}$ 
    \Comment{Do backpropagation}
    
}

\label{algo:pseudo_code}
\end{algorithm}

\section{Proof of Eqn. \eqref{eq:jensen_inequality}}

The upper bound of the objective function Eqn.\eqref{eq:objective_noise} can be derived via Jensen's inequality as follows:
\begin{equation}
\begin{split}
    \sum_{i=1}^L - \log p_\omega(\Lambda_i, \tilde{\rho}_{i-1})
    &= - \sum_{i=1}^L \log \int p_\omega(\Lambda_i, \tilde{\rho}_{i-1}, \tilde{\rho}_i) d \tilde{\rho}_i \\
    &= - \sum_{i=1}^L \log \int p_\omega(\Lambda_i, \tilde{\rho}_{i-1}, \tilde{\rho}_i) \frac{q(\tilde{\rho}_i | \tilde{\rho}_{i-1})}{q(\tilde{\rho}_i | \tilde{\rho}_{i-1})} d \tilde{\rho}_i \\
    &= - \sum_{i=1}^L \log \mathbb{E}_{q} \frac{p_\omega(\Lambda_i, \tilde{\rho}_{i-1}, \tilde{\rho}_i)}{q(\tilde{\rho}_i | \tilde{\rho}_{i-1})} \\
    &\leq \mathbb{E}_q\left[
    - \sum_{i=1}^L \log \frac{p_\omega(\Lambda_i, \tilde{\rho}_{i-1}, \tilde{\rho}_i)}{q(\tilde{\rho}_i | \tilde{\rho}_{i-1})}
    \right] := \mathcal{L}
\end{split}
\end{equation}

\section{Proof of Eqn. \eqref{eq:kl_expand}}

\begin{equation}
\begin{split}
    \mathcal{L} &= \mathbb{E}_q\left[
    - \sum_{i=1}^L \log \frac{p_\omega(\Lambda_i, \tilde{\rho}_{i-1}, \tilde{\rho}_i)}{q(\tilde{\rho}_i | \tilde{\rho}_{i-1})}
    \right] \\
    &= \mathbb{E}_q\left[
    - \sum_{i=1}^L \log \frac{p_\omega(\tilde{\rho}_{i-1} | \Lambda_i, \tilde{\rho}_i)}{q(\tilde{\rho}_{i-1} | \tilde{\rho}_i)}
    p_\omega(\Lambda_i, \tilde{\rho}_i)
    \frac{q(\tilde{\rho}_{i-1})}{q(\tilde{\rho}_i)}
    \right] \\
    &= \mathbb{E}_q\left[
    - \sum_{i=1}^L \log \frac{p_\omega(\tilde{\rho}_{i-1} | \Lambda_i, \tilde{\rho}_i)}{q(\tilde{\rho}_{i-1} | \tilde{\rho}_i)}
    - \sum_{i=1}^L \log p_\omega(\Lambda_i, \tilde{\rho}_i)
    - \log \frac{q(\tilde{\rho}_{0})}{q(\tilde{\rho}_L)}
    \right]\\
    &= \mathbb{E}_q\left[
    \sum_{i=1}^L D_{KL}\left(
    p_\omega(\tilde{\rho}_{i-1} | \Lambda_i, \tilde{\rho}_i) || q(\tilde{\rho}_{i-1} | \tilde{\rho}_i)
    \right)
    - \sum_{i=1}^L \log p_\omega(\Lambda_i, \tilde{\rho}_i)
    - \log \frac{q(\tilde{\rho}_{0})}{q(\tilde{\rho}_L)}
    \right]
\end{split}
\end{equation}
It should be noted that the fraction $\frac{q(\tilde{\rho}_0)}{q(\tilde{\rho}_L)}$ could be considered as constants.
Thus, it should be ignored during the optimization process.

\section{Parameterized Layer Diagram}

\begin{figure}[h]
    \centering
    \includegraphics[width=0.6\linewidth]{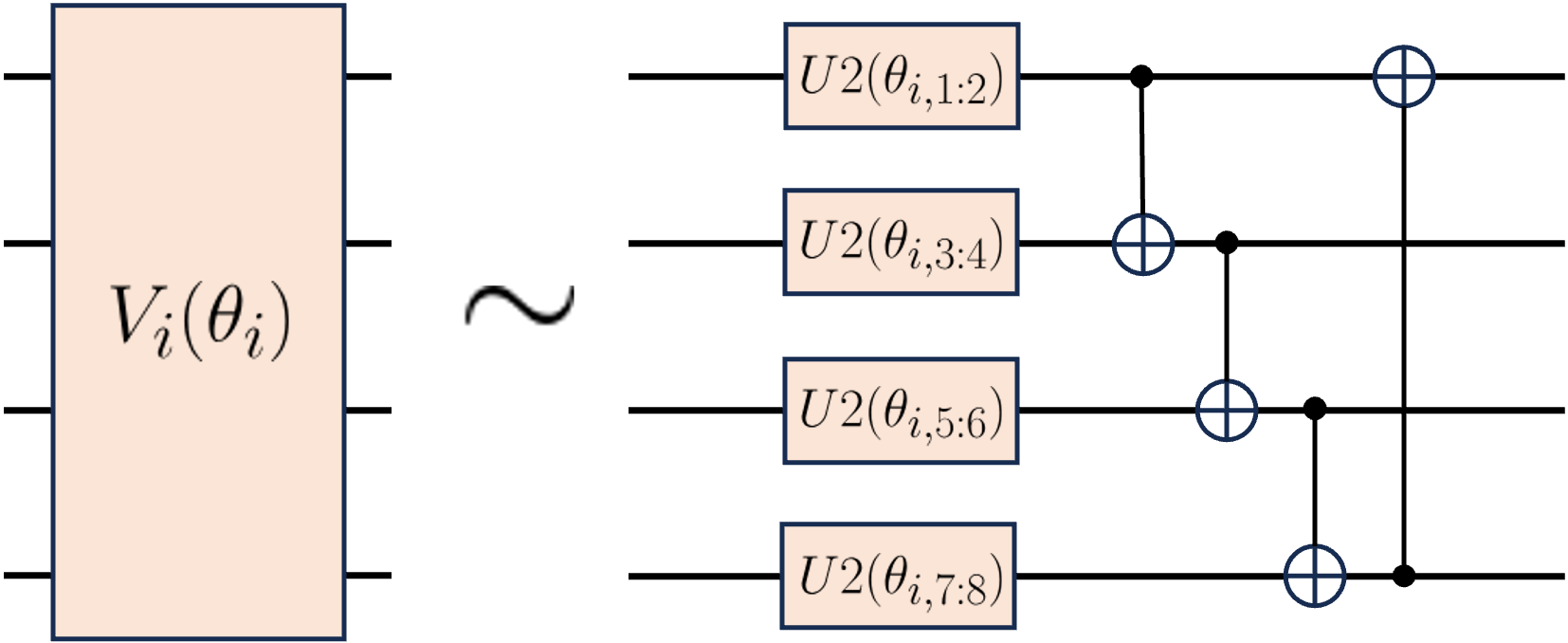}
    \caption{\textbf{The parameterized layer diagram.} The learnable parameters $\theta_{i, j:j+1}$ contain two parameters for X-axis and Y-axis rotation.}
    \label{fig:parameterized_layer_diagram}
\end{figure}

Fig. \ref{fig:parameterized_layer_diagram} shows the diagram of the actual $i$-th parameterized layer in the U2 + CNOT design. The learnable parameters $\theta_{i, j:j+1}$ contain two parameters for X-axis and Y-axis rotation. The RX + CNOT and U3 + CNOT designs have similar diagrams.

\section{Design of Quantum Noise Mitigation Layer}

\begin{figure}[h]
    \centering
    \includegraphics[width=0.7\linewidth]{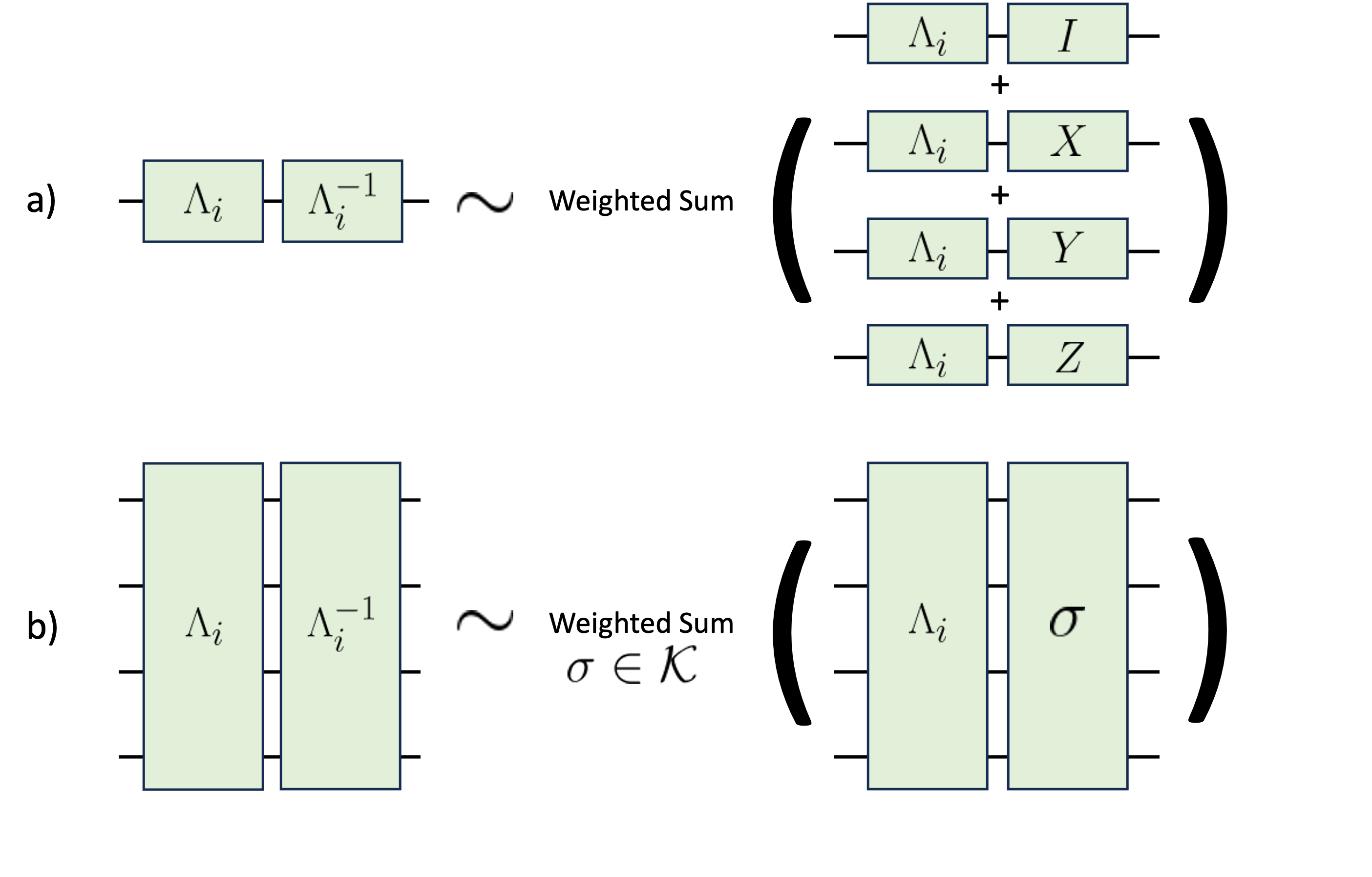}
    \caption{
    \textbf{The quantum noise mitigation module.}
    In the case of 4 qubits, the set of Pauli channels is defined as combinations of Pauli gates $\mathcal{K} = \{\text{IZZZ}, \text{IYYY}, \text{IXYZ}, \text{XYXZ}, \dots\}$.
    }
    \label{fig:noise_mitigation_module}
\end{figure}

\noindent
Fig. \ref{fig:noise_mitigation_module} shows the quantum noise mitigation design diagram in the Pauli noise scenario. 
In the case of 4 qubits, the set of Pauli channels is defined as combinations of Pauli gates $\mathcal{K} = \{\text{IZZZ}, \text{IYYY}, \text{IXYZ}, \text{XYXZ}, \dots\}$.
For each Pauli gate, we define a learnable coefficient. Then, we combine separated measurement results with these coefficients for the final result.

\end{appendices}


\bibliography{sn-bibliography}

\end{document}